\documentclass[aps,preprint]{revtex4}%
\usepackage{amsfonts}
\usepackage{amsmath}
\usepackage{amssymb}%

\begin{document}
\preprint{ }
\title[Relativity in Clifford's Geometric Algebras]{Relativity in Clifford's Geometric Algebras of Space and Spacetime}
\author{William E. Baylis\thanks{email address: baylis@uwindsor.ca}}
\affiliation{Dept. of Physics, University of Windsor, Windsor, ON, Canada N9B 3P4}
\author{Garret Sobczyk\thanks{email address: sobczyk@mail.udlap.mx}}
\affiliation{Dept. de F\'{\i}sica y Matem\'{a}ticas, Universidad de las Am\'{e}ricas,
Cholula, Puebla, M\'{e}xico 72820}

\begin{abstract}
Of the various formalisms developed to treat relativistic phenomena, those
based on Clifford's geometric algebra are especially well adapted for clear
geometric interpretations and computational efficiency. Here we study
relationships between formulations of special relativity in the spacetime
algebra (STA) $C\!\ell_{1,3}$ of the underlying Minkowski vector space, and in
the algebra of physical space (APS) $C\!\ell_{3}$. STA lends itself to an
absolute formulation of relativity, in which paths, fields, and other physical
properties have observer-independent representations. Descriptions in APS are
related by a one-to-one mapping of elements from APS to the even subalgebra
STA$^{+}$ of STA. With this mapping, reversion in APS corresponds to hermitian
conjugation in STA. The elements of STA$^{+}$ are all that is needed to
calculate physically measurable quantities (called \emph{measurables}) because
only they entail the observer dependence inherent in any physical measurement.
As a consequence, every relativistic physical process that can be modeled in
STA also has a representation in APS, and \emph{vice versa.} In the presence
of two or more inertial observers, two versions of APS present themselves. In
the \emph{absolute} version, both the mapping to STA$^{+}$ and hermitian
conjugation are observer dependent, and the proper basis vectors of any
observer are persistent vectors that sweep out timelike planes in spacetime.
To compare measurements by different inertial observers in APS, we express
them in the proper algebraic basis of a single observer. This leads to the
\emph{relative }version of APS, which can be related to STA by assigning every
inertial observer in STA to a single absolute frame in STA. The equivalence of
inertial observers makes this permissible. The mapping and hermitian
conjugation are then the same for all observers. Relative APS gives a
covariant representation of relativistic physics with spacetime multivectors
represented by multiparavectors in APS. We relate the two versions of APS as
consistent models within the same algebra.

\end{abstract}
\maketitle

\section{Introduction}

Clifford's geometric algebra offers a powerful unifying language for the study
of physics
\cite{Hes66,Hes84,Hes87,Hes03,Bay96,Loun97,Ablam03,Snygg97,GuSp97,Dor03}.
Relativistic problems can be treated efficiently both in the spacetime algebra
(STA) \cite{Hes74,Hes03}, the Clifford algebra $C\!\ell_{1,3}$ of Minkowski
spacetime, and in the algebra of physical space (APS)
\cite{Bay99,Bay89,Bay80,Hes87}, the Clifford algebra $C\!\ell_{3}$ of
three-dimensional Euclidean physical space. As real algebras, STA operates in
a linear space of sixteen dimensions, whereas the linear space of APS has
eight dimensions. APS can also be viewed as the 4-dimensional algebra of
complex paravectors \cite{Sob81}. It is also isomorphic to complex
quaternions, which have along history of applications in
relativity,\cite{Silberstein1912,Conway1912} but whose geometrical
interpretation is less obvious.

The connection between STA and APS highlights an important relation, only
rarely explicitly expressed, between the basis vectors in Newtonian mechanics
and those of special relativity. Newtonian basis vectors are usually viewed as
unit displacement vectors in physical space that \emph{persist} in time.
Relativistically, such persistent vectors are associated with
\textquotedblleft timelike\textquotedblright\ \emph{bivectors} (bivectors with
positive squares, see below) representing planes that are swept out \emph{in
spacetime} by a given spatial direction with the passage of time. Under
boosts, timelike bivectors transform into spacetime bivectors that have picked
up spatial bivector parts (with negative squares), and the corresponding
Newtonian basis vectors transform into a mixture of a spatial vector and a
spatial bivector (a plane).

The formulation of relativity in STA can be characterized as \emph{absolute,}
in that physical paths, fields, and other properties of objects are expressed
independent of any observer. The formulation in APS can be either
\emph{absolute} \cite{Sob81} or \emph{relative} \cite{Bay99,Bay89,Bay80}. In
both APS versions, the relationship between APS and STA$^{+}$ is expressed as
an algebra isomorphism together with an operation called \emph{hermitian
conjugation}. Hermitian conjugation, identified with \emph{reversion} in APS,
separates elements of APS into real and imaginary parts, the real part
consisting of scalars and vectors (elements of grade 0 and 1), and the
imaginary part consisting of pseudoscalars (trivectors) and bivectors
(elements of grade 3 and 2).

If we consider only one inertial observer, the two APS versions are
equivalent. In the \emph{absolute} APS approach, one posits a distinct
absolute frame and hermitian conjugation for each observer, so that the
reality and the grading of the elements of APS are observer dependent. The
\emph{relative} APS approach can be related to STA by assuming that all
observers use a single absolute inertial frame in STA to form their
\emph{proper basis}. Since all inertial frames in STA are equivalent, each
inertial observer can identify her frame with the chosen inertial frame. In
this way, the same hermitian conjugation is used for all observers. The
relative APS formulation admits a covariant formulation, of critical
importance to physicists, in which real paravectors are naturally associated
with spacetime vectors and higher paravector grades represent other covariant
geometrical objects in spacetime. While its formulation and
justification\cite{Bay99,Bay96} does not depend on STA, its connection to
STA$^{+}$ clearly shows the relationship between the conceptually different
relative and absolute approaches.

In the following sections, we first review APS and STA, and then construct APS
from the even subalgebra STA$^{+}$ of STA. The formulation of APS is obtained
initially for a single observer. This part is largely a review of previous
work\cite{Hes74} but is included here for completeness. The connections become
more convoluted when additional observers are added, as we investigate first
in the absolute version of APS. We then see how, through the relation of
measurable coefficients of covariant spacetime elements, we are led to the
relative version. We can recover STA from APS by formulating relations in the
proper basis of a single inertial observer in APS assigned to the absolute
frame of STA. From these relationships we conclude that every physical process
formulated in STA can equally well be described in APS.

\section{Review of APS}

We present here a brief summary of the use of APS to model relativistic
phenomena. More details can be found elsewhere \cite{Bay96,Bay99}. The
elements of APS are the real vectors $\mathbf{u,v,w}$ of physical space
$\mathbb{R}^{3}$ and all their sums and products $\mathbf{uu,uv+uvw,\ldots}$.
The elements of APS form an associative algebra under addition and
multiplication. We will see later in the formulation of APS as STA$^{+}$ that
the concept of a \emph{vector} is itself, implicitly, \emph{relative }to an observer.

The square of any vector $\mathbf{u}\in APS$ is defined as its length squared
:%
\begin{equation}
\mathbf{u^{2}}\equiv\mathbf{uu}=\mathbf{u\cdot u~,} \label{axiom}%
\end{equation}
where $\mathbf{u}\cdot\mathbf{u}$ is the usual inner product. This axiom,
together with the usual rules for adding and multiplying square matrices,
determines the entire algebra. If we put $\mathbf{u}=\mathbf{v+w,}$ the axiom
implies%
\begin{equation}
\mathbf{vw+wv}=2\mathbf{v\cdot w~.} \label{corollary}%
\end{equation}
Evidently the algebra is not commutative and, in particular, the product of
perpendicular vectors is anticommutative $\mathbf{uv}=-\mathbf{vu}$. It is
called a \emph{bivector} and has a geometric interpretation as the oriented
plane containing $\mathbf{u}$ and $\mathbf{v.}$

Let $\left\{  \mathbf{e}_{1},\mathbf{e}_{2},\mathbf{e}_{3}\right\}  $ be an
orthonormal basis of $\mathbb{R}^{3}$. The corollary (\ref{corollary}) implies
that
\begin{equation}
\mathbf{e}_{j}\mathbf{e}_{k}+\mathbf{e}_{k}\mathbf{e}_{j}=2\delta_{jk}~,
\label{APSaxiom}%
\end{equation}
where the Kronecker delta $\delta_{jk}$ gives the Euclidean metric of physical
space. For example, $\mathbf{e}_{1}^{2}=1$ and $\mathbf{e}_{1}\mathbf{e}%
_{2}=-\mathbf{e}_{2}\mathbf{e}_{1}.$ The bivector $\mathbf{e}_{1}%
\mathbf{e}_{2}$, a multivector of grade 2, represents a directed area in the
plane of the vectors. Its \textquotedblleft direction\textquotedblright%
\ corresponds to circulation in the plane: if the circulation is reversed, the
sign of the bivector is reversed. The bivector replaces the (Gibbs-Heaviside)
\emph{vector cross product }of polar vectors, but unlike the usual cross
product, it is \emph{intrinsic} to the plane and can be applied to planes in
spaces of any number of dimensions.

The unit bivector $\mathbf{e}_{1}\mathbf{e}_{2}$ squares to
\begin{equation}
(\mathbf{e}_{1}\mathbf{e}_{2})^{2}=-\mathbf{e}_{2}\mathbf{e}_{1}\mathbf{e}%
_{1}\mathbf{e}_{2}=-1
\end{equation}
and generates rotations and reflections in the plane of $\mathbf{e}_{1}$ and
$\mathbf{e}_{2}.$ A general vector $\mathbf{v,}$ with components both in the
plane and perpendicular to it, is rotated through the angle $\phi$ in the
$\mathbf{e}_{1}\mathbf{e}_{2}$ plane by
\begin{equation}
\mathbf{v}\rightarrow R\mathbf{v}R^{\dag}, \label{rotation}%
\end{equation}
where the \emph{rotors\ }$R,R^{\dag}$ are
\begin{align}
R  &  =\exp\left(  -\mathbf{e}_{1}\mathbf{e}_{2}\phi/2\right)  =\cos\frac
{\phi}{2}-\mathbf{e}_{1}\mathbf{e}_{2}\sin\frac{\phi}{2}\\
R^{\dag}  &  =\cos\frac{\phi}{2}-\left(  \mathbf{e}_{1}\mathbf{e}_{2}\right)
^{\dag}\sin\frac{\phi}{2}=\cos\frac{\phi}{2}-\mathbf{e}_{2}\mathbf{e}_{1}%
\sin\frac{\phi}{2}=R^{-1}~.
\end{align}

The dagger $\dag$ used above denotes the conjugation of \emph{reversion,}
which reverses the order of vectors in all products. Thus, for any vector
$\mathbf{v,\ v}^{\dag}=\mathbf{v,}$ and the reversion of a product, say $AB,$
of arbitrary elements can be found from $\left(  AB\right)  ^{\dag}=B^{\dag
}A^{\dag}.$ (A tilde \symbol{126}\ is used in STA to indicate reversion, but
when we associate APS with the even subalgebra STA$^{+}$ of STA we require
distinct symbols.) In spaces of definite metric such as Euclidean spaces, one
commonly represents the basis vectors by hermitian matrices. The dagger then
corresponds to hermitian conjugation and can be used to split elements into
\textquotedblleft real\textquotedblright\ (hermitian) and \textquotedblleft
imaginary\textquotedblright\ (antihermitian) parts:%
\begin{align}
A  &  =\left\langle A\right\rangle _{\Re}+\left\langle A\right\rangle _{\Im}\\
\left\langle A\right\rangle _{\Re}  &  =\frac{A+A^{\dag}}{2},\ \left\langle
A\right\rangle _{\Im}=\frac{A-A^{\dag}}{2}~.
\end{align}

The \emph{standard basis }of APS over the reals can be specified by
\[
\{1,\mathbf{e}_{1},\mathbf{e}_{2},\mathbf{e}_{3},\mathbf{e}_{2}\mathbf{e}%
_{3},\mathbf{e}_{3}\mathbf{e}_{1},\mathbf{e}_{1}\mathbf{e}_{2},\mathbf{e}%
_{1}\mathbf{e}_{2}\mathbf{e}_{3}\},
\]
where the \emph{trivector} $\mathbf{e}_{1}\mathbf{e}_{2}\mathbf{e}_{3}$
squares to $-1.$ In APS, $\mathbf{e}_{1}\mathbf{e}_{2}\mathbf{e}_{3}$ is the
\emph{volume element}, also known as the \emph{unit pseudoscalar.} It commutes
with every vector and hence with all elements and can therefore be identified
with the unit imaginary:%
\begin{equation}
\mathbf{e}_{1}\mathbf{e}_{2}\mathbf{e}_{3}=i~.
\end{equation}
The \emph{center} of APS (the part that commutes with all elements) is spanned
by $\left\{  1,i\right\}  $ and is identified with the complex field.
Bivectors can be identified as imaginary vectors (pseudovectors) in APS. For
example, $\mathbf{e}_{1}\mathbf{e}_{2}=\mathbf{e}_{1}\mathbf{e}_{2}%
\mathbf{e}_{3}\mathbf{e}_{3}=i\mathbf{e}_{3}.$ We can now take the set
\[
\{1,\mathbf{e}_{1},\mathbf{e}_{2},\mathbf{e}_{3}\}
\]
as the standard basis of APS over the complex scalars.

The sum of a real scalar and a real vector is called a \emph{paravector.} A
typical paravector $p$ can be expanded%
\begin{equation}
p=p^{0}+\mathbf{p}=p^{\mu}\mathbf{e}_{\mu},
\end{equation}
where, for notational convenience in using the compact Einstein summation
convention, we put $\mathbf{e}_{0}=1.$ The convention is that repeated
lower-case Greek indices are summed over $0,1,2,3,$ whereas repeated
lower-case Latin indices are summed over the spatial values $1,2,3.$ Every
element in APS can be expressed as a complex paravector. Reversion (the dagger
conjugation) complex-conjugates the complex coefficients and thus changes the
sign of the pseudoscalar and pseudovector parts. Real paravector space is a
four-dimensional linear space spanned by the basis $\left\{  \mathbf{e}%
_{0},\mathbf{e}_{1},\mathbf{e}_{2},\mathbf{e}_{3}\right\}  $ over the reals.

Just as for complex numbers, a natural quadratic form in paravector space is
given by
\begin{equation}
Q\left(  p\right)  =p\bar{p}, \label{Qform}%
\end{equation}
where $\bar{p}=p^{0}-\mathbf{p}$ is called the \emph{Clifford conjugate }of
$p.$ Clifford conjugation is extended to general elements $A,B,$ of APS as an
antiautomorphism: $\overline{AB}=\bar{B}\bar{A}.$ It conveniently splits
elements into \emph{scalarlike} (S) and \emph{vectorlike} (V) parts:%
\begin{align}
A  &  =\left\langle A\right\rangle _{S}+\left\langle A\right\rangle _{V}\\
\left\langle A\right\rangle _{S}  &  =\frac{A+\bar{A}}{2},\ \left\langle
A\right\rangle _{V}=\frac{A-\bar{A}}{2}~.
\end{align}
The quadratic form (\ref{Qform}) is scalarlike. Through it, paravector space
inherits from the Euclidean metric of the underlying space of spatial vectors
an inner product with the Minkowski spacetime metric:
\begin{equation}
\left(  p,q\right)  =\left\langle p\bar{q}\right\rangle _{S}=\frac{p\bar
{q}+q\bar{p}}{2}=p^{\mu}q^{\nu}\eta_{\mu\nu}%
\end{equation}
with the metric tensor
\begin{equation}
\eta_{\mu\nu}=\left\langle \mathbf{e}_{\mu}\mathbf{\bar{e}}_{\nu}\right\rangle
_{S}=\left\{
\begin{array}
[c]{rl}%
1, & \mu=\nu=0\\
-1, & \mu=\nu=1,2,3\\
0, & \mu\neq\nu
\end{array}
\right.  . \label{Minkowski}%
\end{equation}

The appearance of the Minkowski spacetime metric suggests the use of real
paravectors to model vectors in four-dimensional spacetime. Oriented planes in
spacetime are then modeled by \emph{biparavectors }\cite{Bay99} such as
$\left\langle p\bar{q}\right\rangle _{V}=p^{\mu}q^{\nu}\left\langle
\mathbf{e}_{\mu}\mathbf{\bar{e}}_{\nu}\right\rangle _{V},$ which represents
the plane containing all linear combinations of the real paravectors $p$ and
$q,$ and which generally has both real (vector) and imaginary (bivector)
parts. Rotations in paravector space are generated by biparavectors and leave
the quadratic form (and hence inner products) of paravectors invariant. They
represent physical Lorentz transformations. Rotations of the paravector $p$ in
a single spacetime plane have the form%
\begin{equation}
p\rightarrow LpL^{\dag}, \label{Lrotp}%
\end{equation}
where $L$ is a \emph{Lorentz rotor} of the form $L=\exp\left(  \mathbf{W}%
/2\right)  $ and $\mathbf{W}$ is a biparavector for the plane of rotation. In
the special case that $\mathbf{W}$ is imaginary, $L$ is a spatial rotation,
and in the special case that it is real, $L$ is a boost (or hyperbolic
rotation). More generally, any Lorentz rotor $L$ can be factored into the
product $L=BR$ of a spatial rotation $R$ and a boost $B.$ This formulation is
pursued below in the relative formulation of APS.

\section{Review of STA}

STA, introduced by Hestenes \cite{Hes66,Hes74}, is the geometric algebra
$C\!\ell_{1,3}$ of Minkowski spacetime. Minkowski spacetime has a
pseudo-Euclidean metric that highlights the intrinsic difference between time
and space. The aspects of STA presented here are those needed in our
discussion below. Since APS can be equated to the even subalgebra of STA,
particular attention is paid to this subalgebra. Each abstract inertial frame
of STA consists of a constant 4-dimensional orthonormal vector basis $\left\{
\gamma_{0},\gamma_{1},\gamma_{2},\gamma_{3}\right\}  \equiv\left\{
\gamma_{\mu}\right\}  $ satisfying%
\begin{equation}
\gamma_{\mu}\gamma_{\nu}+\gamma_{\nu}\gamma_{\mu}=2\eta_{\mu\nu},
\label{STAaxiom}%
\end{equation}
where $\eta_{\mu\nu}$ are elements (\ref{Minkowski}) of the Minkowski
spacetime metric tensor.

The \emph{history} (worldline) of an idealized point particle $P$ is a
timelike curve $r^{P}\left(  \tau^{P}\right)  $ giving its position in STA as
a function of a scalar parameter $\tau^{P}$, which we take to be its proper
time. The tangent vector $u^{P}=dr^{P}/d\tau^{P}$ is its \emph{proper velocity
}(in units with $c=1$). In a \emph{commoving frame }(an inertial frame
instantaneously moving with the observer) $\left\{  \gamma_{\mu}^{P}\right\}
$ of the particle, the displacement has only a time component:%
\[
dr^{P}=\gamma_{0}^{P}d\tau^{P},
\]
and the commoving unit time axis $\gamma_{0}^{P}$ is thus seen to be
coincident with the proper velocity: $u^{P}=\gamma_{0}^{P}~.$

The spacetime curve $r^{P}\left(  \tau^{P}\right)  $ and its tangent vector
$u^{P}\ $are \emph{abstract }and do not, by themselves, determine physically
measurable values. They are \emph{independent of the observer} and therefore
unchanged by \emph{passive} transformations (transformations of the observer).
For this reason Hestenes\cite{Hes03} calls such spacetime quantities
\emph{invariant}. However, they do transform under \emph{active }Lorentz
rotations, and to avoid possible confusion with \emph{Lorentz scalars}, which
are invariant under both passive and active transformations, we prefer to call
them \emph{absolute}. Physically measurable quantities
(\emph{\textquotedblleft measurables\textquotedblright}), on the other hand,
are either Lorentz invariants or are derived from vector components
\emph{relative }to the observer. As we show below, these are determined by
even elements of STA.

An \emph{inertial observer}, say Alice, can be idealized as a congruence of
parallel histories $r^{A}\left(  \tau^{A}\right)  ,$ all with the same given
proper velocity $u^{A}=dr^{A}/d\tau^{A}$ with $du^{A}/d\tau^{A}=0$ and a
constant commoving frame $\left\{  \gamma_{\mu}^{A}\right\}  $ with
$\gamma_{0}^{A}=u^{A}.$ When Alice measures physical quantities represented by
geometric objects such as vectors, she normally expresses them in her frame
$\left\{  \gamma_{\mu}^{A}\right\}  $ of spacetime basis vectors. The
$\gamma_{\mu}^{A}$ are abstract, and it is the scalar coefficients of the
expansion that constitute the measurables for Alice. A different observer, say
Bob, has his own frame $\left\{  \gamma_{\mu}^{B}\right\}  $ with $\gamma
_{0}^{B}=u^{B}$ that he normally uses for measurements. We assume that the
handedness of the two frames is the same:%
\[
\gamma_{0}^{A}\gamma_{1}^{A}\gamma_{2}^{A}\gamma_{3}^{A}=\gamma_{0}^{B}%
\gamma_{1}^{B}\gamma_{2}^{B}\gamma_{3}^{B}\equiv\mathbf{I}~,
\]
where the pseudoscalar $\mathbf{I}$ of STA anticommutes with vectors.

Any two inertial frames of the same handedness are related by a \emph{Lorentz
rotation,} also known as a restricted (proper orthochronous) Lorentz
transformation, and is specified by a rotor $L.$ Every Lorentz rotation can be
expressed as the product of a \emph{spatial rotation }and a \emph{boost }(a
velocity transformation). There is in particular a rotor $L$ that relates
Alice's frame to Bob's:%
\begin{equation}
\gamma_{\mu}^{B}=L\gamma_{\mu}^{A}\tilde{L}~, \label{ABtransf}%
\end{equation}
where $\symbol{126}$ indicates reversion in STA. Lorentz rotors are
\emph{unimodular}:
\begin{equation}
L\tilde{L}=1, \label{STAunimod}%
\end{equation}
and consequently all products of spacetime vectors transform in the same way
in STA. For $\mu=0$, the transformation (\ref{ABtransf}) relates the proper
velocities%
\begin{equation}
\gamma_{0}^{B}=u^{B}=L\gamma_{0}^{A}\tilde{L}=Lu^{A}\tilde{L}~.
\end{equation}

It is common to define hermitian conjugation of any element $K$ for Alice by%
\begin{equation}
K^{\dag A}=\gamma_{0}^{A}\tilde{K}\gamma_{0}^{A}. \label{STAdag}%
\end{equation}
The conjugation symbol $\dag A$ is used here instead of the more usual $\dag$
to underscore its dependence on the observer's frame. In terms of the
hermitian conjugate, the proper velocities are related by%
\begin{align}
\gamma_{0}^{B}  &  =u^{B}=LL^{\dag A}\gamma_{0}^{A}=LL^{\dag A}u^{A}%
\nonumber\\
LL^{\dag A}  &  =\gamma_{0}^{B}\gamma_{0}^{A}=u^{B}u^{A}, \label{Lorentzrot}%
\end{align}
where we noted from (\ref{STAaxiom}) that $\left(  \gamma_{0}^{A}\right)
^{2}=1.$ If the frames of Alice and Bob are related by a pure spatial
rotation, their time axes and hence proper velocities $u^{B}$ and $u^{A}$ are
equal and $L^{\dag A}=\tilde{L}=L^{\dag B}.$ On the other hand, if their
frames are related by a pure boost, then $L^{\dag A}=L=L^{\dag B}$ and
$L^{2}=\gamma_{0}^{B}\gamma_{0}^{A}=u^{B}u^{A}.$ In the most general case, the
Lorentz rotor $L$ is the product of boost and spatial-rotation rotors, and
$LL^{\dag A}$ is simply the square of the boost rotor.

A boost along the $\gamma_{1}^{A}$ direction is a Lorentz rotation generated
by the STA bivector $\gamma_{10}^{A}\equiv\gamma_{1}^{A}\gamma_{0}^{A},$ and
it has the form $\exp\left(  w\gamma_{10}^{A}/2\right)  ,$ where the scalar
parameter $w$ is called the rapidity of the boost. We can expand the
exponential to get for the rotor $L$%
\begin{equation}
L^{2}=\exp\left(  w\gamma_{10}^{A}\right)  =\gamma\left(  1+\mathbf{v}\right)
=u^{B}u^{A},
\end{equation}
where $\gamma\equiv\cosh w$ is the Lorentz time dilation factor between the
observers, and $\mathbf{v}=\gamma_{10}^{A}\tanh w$ gives the relative
coordinate velocity of Bob with respect to Alice. The coordinate velocity of
Alice as seen by Bob is $-\mathbf{v.}$ The plane of the Lorentz rotation is
itself invariant under $L$: $\gamma_{10}^{B}=L\gamma_{10}^{A}\tilde{L}%
=\gamma_{10}^{A},$ and $\gamma$ and $\left\vert \mathbf{v}\right\vert $ are
measurables for both Alice and Bob.

We want to compare measurements made by Alice and Bob of an absolute spacetime
vector, which for concreteness we take to be the spacetime position $r.$
(Since $r$ depends on the origin of coordinates, in order that this transform
under fixed $L$ as a spacetime vector, we must assume that the origins of
Alice's and Bob's frames coincide. Alternatively, we could replace $r$ by an
affine spacetime vector.) Thus, $r$ might be one point on the history of a
point particle or some other event. It can be expanded in the basis vectors of
any inertial frame, for example in the frame $\left\{  \gamma_{\mu}%
^{A}\right\}  $ of Alice:%
\begin{equation}
r=r_{A}^{\mu}\gamma_{\mu}^{A}~.
\end{equation}

As emphasized above, the spacetime vectors $r$ and $\gamma_{\nu}^{A}$ are
abstract and not directly measurable; it is the scalar coefficients
$r_{A}^{\mu}$ that Alice can measure. The time component of $r$ measured by
Alice is%
\begin{equation}
t_{A}\equiv r_{A}^{0}=r\cdot\gamma_{0}^{A}~.
\end{equation}
More generally, the spacetime vector $r$ \emph{relative to Alice }is given by
the expansion%
\begin{equation}
r\gamma_{0}^{A}=r_{A}^{\mu}\gamma_{\mu}^{A}\gamma_{0}^{A}\equiv r_{A}%
=t_{A}+\mathbf{r}_{A}~, \label{ST}%
\end{equation}
where the spacetime bivector $\mathbf{r}_{A}=r\wedge\gamma_{0}^{A}=r_{A}%
^{k}\gamma_{k0}^{A}\in$ STA$^{+},$ with $\gamma_{k0}^{A}\equiv\gamma_{k}%
^{A}\gamma_{0}^{A},$ is interpreted in APS as the spatial \emph{position
vector} from the origin to the particle\emph{ }that Alice measures. The even
element $r_{A}=r\gamma_{0}^{A}$ is called the \emph{relative position,} and
its expansion (\ref{ST}) is called a \emph{space-time }split \emph{
}\cite{Hes74,Hes03}.

Analogous relations can be written for Bob's measurables. The relative
positions $r_{A}=r\gamma_{0}^{A}$ and $r_{B}=r\gamma_{0}^{B}$ are related by
the \emph{passive} Lorentz rotation%
\begin{align}
r_{B}  &  =r\gamma_{0}^{B}=rL\gamma_{0}^{A}\tilde{L}=r\gamma_{0}^{A}\tilde
{L}^{\dag A}\tilde{L}\nonumber\\
&  =r_{A}\tilde{L}^{\dag A}\tilde{L}. \label{passive}%
\end{align}
The inverse transformation can be written%
\begin{equation}
r_{A}=r_{B}LL^{\dag A}=r_{B}L^{\dag B}L~,
\end{equation}
where $L^{\dag B}=\gamma_{0}^{B}\tilde{L}\gamma_{0}^{B}=L\gamma_{0}^{A}%
\tilde{L}\tilde{L}L\gamma_{0}^{A}\tilde{L}=LL^{\dag A}\tilde{L}~.$

A different form of transformation results by boosting the event $r$ for a
given observer. This is an example of an \emph{active} transformation, in
which the observer is fixed and the event is transformed. Let%
\begin{equation}
r^{\prime}=Lr\tilde{L}.
\end{equation}
This boost changes the relative position with respect to Alice according to%
\begin{align}
r_{A}^{\prime}  &  =r^{\prime}\gamma_{0}^{A}=r_{A}^{\prime\mu}\gamma_{\mu}%
^{A}\gamma_{0}^{A}=Lr\tilde{L}\gamma_{0}^{A}=Lr\gamma_{0}^{A}L^{\dag
A}\nonumber\\
&  =Lr_{A}L^{\dag A}~. \label{active}%
\end{align}
If both of these transformations are performed together (the order is not
important when the same rotor is used for both) we obtain the relative
spacetime position of the transformed event with respect to Bob:%
\begin{align}
r_{B}^{\prime}  &  =r^{\prime}\gamma_{0}^{B}=Lr\gamma_{0}^{A}\tilde{L}%
=Lr_{A}^{\mu}\gamma_{\mu}^{A}\gamma_{0}^{A}\tilde{L}\nonumber\\
&  =Lr_{A}\tilde{L}~. \label{both}%
\end{align}

The three transformations of measurables (\ref{passive}), (\ref{active}),
(\ref{both}) have distinct forms. In the transformation (\ref{both}), in which
both the observer and the event are boosted by the same amount, the
unimodularity of $L$ (\ref{STAunimod}) implies that the time measurements are
the same: $t_{B}^{\prime}=t_{A},$ but the positions generally differ%
\begin{equation}
\mathbf{r}_{B}^{\prime}=r_{B}^{\prime k}\gamma_{k0}^{B}=L\mathbf{r}_{A}%
\tilde{L}=Lr_{A}^{j}\gamma_{j0}^{A}\tilde{L}~.
\end{equation}
How are the three transformations to be interpreted physically?

\subsection{Measurables in STA}

To understand the \emph{physical interpretation} of the above equations, it is
important to note that all of the measurables have been expressed in terms of
elements of STA$^{+}$, the even subalgebra of STA, spanned by the basis
$\left\{  1,\gamma_{\mu\nu},I\right\}  ,~$where the basis bivectors in STA are
$\gamma_{\mu\nu}\equiv\frac{1}{2}\left(  \gamma_{\mu}\gamma_{\nu}-\gamma_{\nu
}\gamma_{\mu}\right)  $ with $0\leq\mu<\nu\leq3$. The reason for this is that
every physical measurement of a vector is \emph{relative} and involves
\emph{two }absolute spacetime vectors: the absolute vector of the event or
property being measured and a basis vector of the reference frame used by the
observer. The results of the measurement depend on the orientation and motion
of one with respect to the other.

Of course it is most common for an observer to employ a reference frame at
relative rest. Measurements can then be expressed in STA$^{+}$ in the
observer-dependent \emph{proper basis}
\begin{equation}
\{\boldsymbol{\sigma}_{\mu}^{A}=\gamma_{\mu}^{A}\gamma_{0}^{A}\},\ \left\{
\boldsymbol{\sigma}_{\mu}^{B}=\gamma_{\mu}^{B}\gamma_{0}^{B}\right\}  .
\label{proper}%
\end{equation}
The bases are called \emph{proper} because they are based on the absolute
frames that are at rest with respect to the observer. The basis element
representing the time axis in the relative frame is in each case unity:%
\[
\boldsymbol{\sigma}_{0}^{A}=1=\boldsymbol{\sigma}_{0}^{B}%
\]
and the spatial elements are spacetime bivectors $\boldsymbol{\sigma}%
_{k}=\gamma_{k0}=\gamma_{k}\gamma_{0}$ representing planes containing both the
spatial direction $\gamma_{k}$ and the time direction $\gamma_{0}~.$ These are
the planes swept out by the unit spatial vector $\gamma_{k}$ in one unit of
time. They are the spacetime planes representing \emph{persistent} spatial
vectors in physical space. The relation between the relative basis vectors is%
\begin{equation}
\boldsymbol{\sigma}_{\mu}^{B}=L\boldsymbol{\sigma}_{\mu}^{A}\tilde{L}
\label{sigmatransf}%
\end{equation}
for $\mu=0,1,2,3$, which follows directly from $\gamma_{\mu}^{B}=L\gamma_{\mu
}^{A}\tilde{L}$ and $L\tilde{L}=1.$ We will return to the meaning of this
transformation below.

In terms of these proper bases, the three transformations (\ref{passive}%
,\ref{active},\ref{both}) are%
\begin{equation}%
\begin{tabular}
[c]{ll}%
$r_{A}\rightarrow r_{B}=r_{B}^{\mu}\boldsymbol{\sigma}_{\mu}^{B}=r_{A}%
\tilde{L}^{\dag A}\tilde{L}$ & $\text{passive}$\\
$r_{A}\rightarrow r_{A}^{\prime}=r_{A}^{\prime\mu}\boldsymbol{\sigma}_{\mu
}^{A}=Lr_{A}L^{\dag A}$ & $\text{active}$\\
$r_{A}\rightarrow r_{B}^{\prime}=r_{B}^{\prime\mu}\boldsymbol{\sigma}_{\mu
}^{B}=Lr_{A}\tilde{L}$ & $\text{both.}$%
\end{tabular}
\ \ \ \ \ \ \label{transf1}%
\end{equation}
The measurables for Alice are the scalar coefficients $r_{A}^{\mu}$ and
$r_{A}^{\prime\mu},$ and for Bob, they are $r_{B}^{\mu}$ and $r_{B}^{\prime
\mu}.$ We can relate them using the orthogonality of the basis vectors. For
example, for the passive case,%
\begin{equation}
r_{B}^{\mu}=r_{A}^{\nu}\left\langle \boldsymbol{\sigma}_{\nu}^{A}\tilde
{L}^{\dag A}\tilde{L}\boldsymbol{\sigma}_{\mu}^{B}\right\rangle _{S}%
=r_{A}^{\nu}\left\langle \boldsymbol{\sigma}_{\nu}^{A}\tilde{L}^{\dag
A}\boldsymbol{\sigma}_{\mu}^{A}\tilde{L}\right\rangle _{S}%
\end{equation}
However, the relations are most transparently seen if the transformations
(\ref{transf1}) are expressed in terms of a \emph{proper basis} of a
\emph{single observer}. If we insert the transformation (\ref{sigmatransf})
into (\ref{transf1}), we find%
\begin{equation}%
\begin{tabular}
[c]{ll}%
$r_{B}^{\mu}\boldsymbol{\sigma}_{\mu}^{A}=r_{A}^{\nu}\tilde{L}%
\boldsymbol{\sigma}_{\nu}^{A}\tilde{L}^{\dag A}=\tilde{L}r_{A}\tilde{L}^{\dag
A}$ & $\text{passive}$\\
$r_{A}^{\prime\mu}\boldsymbol{\sigma}_{\mu}^{A}=r_{A}^{\nu}L\boldsymbol{\sigma
}_{\nu}^{A}L^{\dag A}=Lr_{A}L^{\dag A}$ & $\text{active}$\\
$r_{B}^{\prime\mu}\boldsymbol{\sigma}_{\mu}^{A}=r_{A}^{\nu}\boldsymbol{\sigma
}_{\nu}^{A}=r_{A}$ & $\text{both.}$%
\end{tabular}
\label{transf2}%
\end{equation}

Note that the active and passive transformations (\ref{transf2}) have similar
forms but with inverse rotors. They demonstrate in particular that under
either active or passive boosts, time and space components are mixed and time
intervals (and hence clock rates) change. The last relation of (\ref{transf2})
shows that when both the event and the observer are transformed by the same
rotor, the components are unchanged:%
\begin{equation}
r_{B}^{\prime\mu}=r_{A}^{\mu}.
\end{equation}
In other words, the spacetime vector $r^{\prime}$ measured by Bob has the same
components as $r$ measured by Alice.

Notice that $\boldsymbol{\sigma}_{k}^{A}$ and $\boldsymbol{\sigma}_{k}^{B}$,
for $k=1,2,3,$ are timelike bivectors when expressed in the frames $\left\{
\gamma_{\mu}^{A}\right\}  $ and $\left\{  \gamma_{\mu}^{B}\right\}  ,$
respectively, of STA. In the next section, we will reinterpret the
$\boldsymbol{\sigma}_{k}^{A}$ to be the persistent vectors of a single,
absolute inertial frame, and study the resulting geometric algebra, which is
isomorphic to $Cl_{3}$.

\section{APS as STA$^{+}$}

Since APS is isomorphic to the even subalgebra of STA, we can derive the
relativistic formalism for APS by equating the basis vectors $\mathbf{e}_{k}$
of APS\ to the appropriate elements of STA$^{+}.$ For Alice, we have
\begin{equation}
\mathbf{e}_{k}=\boldsymbol{\sigma}_{k}^{A}=\gamma_{k0}^{A}~,\;k=1,2,3.
\label{ident}%
\end{equation}
These three orthonormal vectors satisfy the axiom (\ref{APSaxiom}) and
generate both APS and STA$^{+}$. The identification (\ref{ident}) associates
the three spatial basis vectors of APS with timelike bivectors of STA. This
reinforces the concept of \emph{persistent vectors} in APS that sweep out
timelike planes in STA. Since $\mathbf{\sigma}_{0}^{A}=1,$ we can extend the
identification (\ref{ident}) to Alice's proper paravector basis $\left\{
\mathbf{e}_{\mu}=\boldsymbol{\sigma}_{\mu}^{A}\right\}  $ with%
\begin{equation}
\mathbf{e}_{\mu}=\boldsymbol{\sigma}_{\mu}^{A}=\gamma_{\mu}^{A}\gamma_{0}%
^{A}~,\;\mu=0,1,2,3. \label{e0biv}%
\end{equation}
From this identification, the volume elements of APS and STA are%
\[
i=\mathbf{e}_{1}\mathbf{e}_{2}\mathbf{e}_{3}=\boldsymbol{\sigma}_{1}%
^{A}\boldsymbol{\sigma}_{2}^{A}\boldsymbol{\sigma}_{3}^{A}=\gamma_{10}%
^{A}\gamma_{20}^{A}\gamma_{30}^{A}=\gamma_{0}^{A}\gamma_{1}^{A}\gamma_{2}%
^{A}\gamma_{3}^{A}=\mathbf{I}.
\]
As in STA, the volume element of APS\ squares to $-1,$ but whereas in STA
$\mathbf{I}$ anticommutes with all vectors, in both APS and STA$^{+}$ the
volume element is part of the center of the algebra, that is, it commutes with
all elements.

Clifford conjugation in APS corresponds to reversion in STA$^{+}$%
\begin{align*}
\mathbf{\bar{e}}_{0}  &  =\mathbf{e}_{0}=1=\tilde{1}\\
\mathbf{\bar{e}}_{k}  &  =-\mathbf{e}_{k}=-\gamma_{k0}^{A}=\tilde{\gamma}%
_{k0}^{A}~,
\end{align*}
and it follows that Clifford conjugation of an observer's proper basis is the
same for every observer. Reversion in APS is equivalent to Alice's hermitian
(dagger) conjugation (\ref{STAdag}) in STA$^{+},$ and it follows that the
proper basis vectors $\mathbf{e}_{\mu}$ are real as seen by Alice:%
\[
\mathbf{e}_{\mu}^{\dag}=\boldsymbol{\sigma}_{\mu}^{\dag A}=\gamma_{0}%
^{A}\boldsymbol{\tilde{\sigma}}_{\mu}^{A}\gamma_{0}^{A}=\mathbf{e}_{\mu}.
\]

The equations (\ref{transf1}) for Lorentz rotations in STA$^{+}$ are unchanged
in APS except that, for Alice, $\boldsymbol{\sigma}_{\mu}^{A}$ is replaced by
$\mathbf{e}_{\mu}~.$ The proper basis vectors $\boldsymbol{\sigma}_{\mu}^{B}$
used by Bob, given by (\ref{sigmatransf}), are seen by Alice to be complex.
This can be understood, as mentioned above, as the result of transforming the
timelike planes swept out by persistent vectors in time, and it reflects the
action of the six-parameter group $SL\left(  2,\mathbb{C}\right)  $ of Lorentz
rotors, which mix timelike and spacelike planes in spacetime. Such planes
correspond respectively to vectors and bivectors in APS. The fact that Alice
sees Bob's frame as complex presents no physical problem because the
definition of proper hermitian conjugation (\ref{STAdag}) \emph{depends on the
observer} and ensures that each observer sees her own proper basis vectors as
real \cite{Sob81a}.

Indeed the mapping (\ref{e0biv}) between APS and STA$^{+}\ $is observer
dependent, and while Alice takes the proper basis vectors $\mathbf{e}_{k}$ of
APS to be her timelike bivectors $\boldsymbol{\sigma}_{k}^{A}$ in STA$^{+},$
Bob equates them to his bivectors $\boldsymbol{\sigma}_{k}^{B}$. Just as under
Alice's hermitian conjugation, Bob's proper basis vectors $\boldsymbol{\sigma
}_{k}^{B}$ are complex whereas hers are real, under Bob's conjugation, it is
Alice's $\boldsymbol{\sigma}_{k}^{A}$ that are complex and his are real. By
allowing both the mapping between APS and STA$^{+}$ and hermitian conjugation
to vary with observer, this formulation of APS is thus able to treat all
observers on an equal footing. The proper vector basis used by any observer in
this formulation, like the bivector basis in STA to which it is equated, is
absolute and distinct from the bases used by other observers who are in
relative motion. We therefore refer to this formulation of APS as
\emph{absolute}.

The important relations in relativity, however, are not the way one observer
sees another observer's proper basis, but how the real, scalar values measured
by one observer are related to those measured by another. Such relations are
given by the transformations (\ref{transf2}), and these suggest a relative
formulation of APS.

\subsection{Covariant Multiparavectors}

While any collection of real vector components on the basis $\left\{
\mathbf{e}_{0},\mathbf{e}_{1},\mathbf{e}_{2},\mathbf{e}_{3}\right\}  $ defines
a real paravector, only some such collections transform together under Lorentz
rotations as a spacetime vector. Those that do are said to be \emph{covariant}%
. Furthermore, a spatial vector in APS may be part of a spacetime vector,
whose components transform according to (\ref{transf2}), or it may be part of
a spacetime bivector, whose components transform differently. For example, the
proper acceleration $du/d\tau$ of a point particle is a spatial vector in the
commoving inertial frame, and it transforms as in (\ref{transf2}) above.
However, the electric field $\mathbf{E}$ is also a pure vector, but it
transforms distinctly. In fact, it transforms as a persistent vector and may
be seen by a different observer as having both real vector (electric) and
imaginary vector or bivector (magnetic) parts. Its basis expansion in APS is
simply%
\[
\mathbf{E}=E_{A}^{k}\boldsymbol{\sigma}_{k}^{A}%
\]
for Alice, and naively applying the arguments made for the position vector
$r_{A},$ it would be seen by Bob (using his proper conjugation) to be real and
to pick up a scalar part. It would remain real because Bob's proper
conjugation is different from Alice's, and the difference exactly compensates
the generation of the imaginary part.

This, however, is wrong. There is no meaning to a scalar part of the
electromagnetic field, and Bob really sees a magnetic (imaginary) component to
the field. What is missing is a recognition that the proper acceleration and
the electric-field vector belong to different types of covariant objects and
therefore transform differently under Lorentz rotations. The proper
acceleration is part of a spacetime vector whereas the electric field is part
of a spacetime bivector $\mathbf{F}$ representing the electromagnetic field.

Alice and Bob see different electric and magnetic components of the given
electromagnetic field $\mathbf{F}$ because their hermitian conjugations are
different. Thus,%
\[
\mathbf{F}=\frac{1}{2}(\mathbf{F}+\mathbf{F}^{\dagger A})+\frac{1}%
{2}(\mathbf{F}-\mathbf{F}^{\dagger A})=\mathbf{E}_{A}+i\mathbf{B}_{A}%
\]%
\[
=\frac{1}{2}(\mathbf{F}+\mathbf{F}^{\dagger B})+\frac{1}{2}(\mathbf{F}%
-\mathbf{F}^{\dagger B})=\mathbf{E}_{B}+i\mathbf{B}_{B}%
\]

To relate the measurables, that is the actual field components measured by
Alice and Bob, we expand the covariant spacetime bivector of which it is a
part in an inertial frame in STA and then express the result in a single
proper basis of APS. For Alice the appropriate expansion is%
\begin{align*}
\mathbf{F}  &  =\frac{1}{2}F_{A}^{\mu\nu}\gamma_{\mu\nu}^{A}=\frac{1}{4}%
F_{A}^{\mu\nu}\left(  \gamma_{\mu0}^{A}\gamma_{0\nu}^{A}-\gamma_{\nu0}%
^{A}\gamma_{0\mu}^{A}\right) \\
&  =\frac{1}{4}F_{A}^{\mu\nu}\left(  \boldsymbol{\sigma}_{\mu}^{A}%
\boldsymbol{\tilde{\sigma}}_{v}^{A}-\boldsymbol{\sigma}_{\nu}^{A}%
\boldsymbol{\tilde{\sigma}}_{\mu}^{A}\right)  .
\end{align*}
The same $\mathbf{F}$ can be expanded in Bob's basis $\mathbf{F=}\frac{1}%
{4}F_{B}^{\mu\nu}\left(  \boldsymbol{\sigma}_{\mu}^{B}\boldsymbol{\tilde
{\sigma}}_{\nu}^{B}\mathbf{-}\boldsymbol{\sigma}_{\nu}^{B}\boldsymbol{\tilde
{\sigma}}_{\mu}^{B}\right)  $ so that the relation between the components seen
by Alice and Bob is%
\[
F_{B}^{\mu\nu}\left(  \boldsymbol{\sigma}_{\mu}^{B}\boldsymbol{\tilde{\sigma}%
}_{\nu}^{B}\mathbf{-}\boldsymbol{\sigma}_{\nu}^{B}\boldsymbol{\tilde{\sigma}%
}_{\mu}^{B}\right)  =F_{A}^{\mu\nu}\left(  \boldsymbol{\sigma}_{\mu}%
^{A}\boldsymbol{\tilde{\sigma}}_{v}^{A}-\boldsymbol{\sigma}_{\nu}%
^{A}\boldsymbol{\tilde{\sigma}}_{\mu}^{A}\right)  ,
\]
where $\boldsymbol{\sigma}_{\mu}^{B}\boldsymbol{\tilde{\sigma}}_{\nu}%
^{B}-\boldsymbol{\sigma}_{\nu}^{B}\boldsymbol{\tilde{\sigma}}_{\mu}%
^{B}=L\left(  \boldsymbol{\sigma}_{\mu}^{A}\boldsymbol{\tilde{\sigma}}_{v}%
^{A}-\boldsymbol{\sigma}_{\nu}^{A}\boldsymbol{\tilde{\sigma}}_{\mu}%
^{A}\right)  \tilde{L}~.$ This leads to the relation%
\[
F_{B}^{\mu\nu}\left(  \boldsymbol{\sigma}_{\mu}^{A}\boldsymbol{\tilde{\sigma}%
}_{v}^{A}-\boldsymbol{\sigma}_{\nu}^{A}\boldsymbol{\tilde{\sigma}}_{\mu}%
^{A}\right)  =4\tilde{L}\mathbf{F}L=\tilde{L}F_{A}^{\mu\nu}\left(
\boldsymbol{\sigma}_{\mu}^{A}\boldsymbol{\tilde{\sigma}}_{v}^{A}%
-\boldsymbol{\sigma}_{\nu}^{A}\boldsymbol{\tilde{\sigma}}_{\mu}^{A}\right)  L
\]
under a passive Lorentz rotation. With the identification (\ref{e0biv}), in
APS this becomes%
\begin{equation}
F_{B}^{\mu\nu}\left\langle \mathbf{e}_{\mu}\mathbf{\bar{e}}_{\nu}\right\rangle
_{V}=\tilde{L}F_{A}^{\mu\nu}\left\langle \mathbf{e}_{\mu}\mathbf{\bar{e}}%
_{\nu}\right\rangle _{V}L \label{passivebv}%
\end{equation}

Thus, whereas every observer sees the proper acceleration or (for inertial
observers sharing a common spacetime origin) the position as a real paravector
in APS, the boosted electric field becomes a mixture of real and imaginary
vectors, representing the electric and magnetic fields seen by a different
observer. Vectors transform differently depending on what (if any) type of
covariant object they belong to. The ability to identify the type of covariant
object being transformed is essential in establishing the correct
transformations between the measurables (scalar coefficients) for different
observers, and the use of covariant objects such as spacetime vectors and
bivectors is also important for simplifying relations and bringing out the
geometry and relativistic symmetries of the problem.{} In STA, covariant
objects are generally homogenous $k$-vectors. In APS, homogeneous $k$-vectors
are generally not covariant. Instead, it is the paravectors and
multiparavectors that provide the covariant formulation. Their utility is
based on the relative formulation of APS, discussed in the next Section.

\section{Relative APS}

The formulation of relativity in absolute APS depends on the identification of
the proper paravector basis and hermitian conjugation for each observer. Thus,
Alice uses the basis $\left\{  \boldsymbol{\sigma}_{\mu}^{A}\right\}  ,$ which
is real under her conjugation operator $\dag A,$ whereas Bob uses a different
proper basis $\left\{  \boldsymbol{\sigma}_{\mu}^{B}\right\}  $ with the
corresponding operator $\dag B.$ However, as seen above, to relate Alice's and
Bob's measurables such as the scalar coefficients $r_{A}^{\mu}$ and
$r_{B}^{\mu},$ we use transformations (\ref{transf2}) that entail \emph{only
one} proper basis for both observers. In (\ref{transf2}) we used Alice's
proper basis and conjugation, but we would obtain the same results by using
Bob's proper basis or, in fact, the proper basis of any inertial observer
together with that observer's proper conjugation.

To see this, replace $\boldsymbol{\sigma}_{\mu}^{A}$ by $\tilde{L}%
_{CA}\boldsymbol{\sigma}_{\mu}^{C}L_{CA}$ in order to express the result in
Carol's proper basis $\left\{  \boldsymbol{\sigma}_{\mu}^{C}\right\}  .$ Here,
$L_{CA}$ is the Lorentz rotor for the transformation from Alice to Carol. The
passive transformation in (\ref{transf2}), for example, becomes%
\begin{equation}
r_{A}\rightarrow r_{B}^{\mu}\tilde{L}_{CA}\boldsymbol{\sigma}_{\mu}^{C}%
L_{CA}=r_{A}^{\nu}\tilde{L}\tilde{L}_{CA}\boldsymbol{\sigma}_{\nu}^{C}%
L_{CA}\tilde{L}^{\dag A}\ \label{relative}%
\end{equation}
which since $L_{CA}\tilde{L}_{CA}=1$ and $\gamma_{0}^{C}=L_{CA}\gamma_{0}%
^{A}\tilde{L}_{CA},$ is equivalent to%
\begin{align}
r_{B}^{\mu}\boldsymbol{\sigma}_{\mu}^{C}  &  =r_{A}^{\nu}L_{CA}\tilde{L}%
\tilde{L}_{CA}\boldsymbol{\sigma}_{\nu}^{C}L_{CA}\tilde{L}^{\dag A}\tilde
{L}_{CA}\nonumber\\
&  =r_{A}^{\nu}L_{CA}\tilde{L}\tilde{L}_{CA}\boldsymbol{\sigma}_{\nu}%
^{C}\tilde{L}_{AC}\gamma_{0}^{A}L\gamma_{0}^{A}\tilde{L}_{CA}\nonumber\\
&  =r_{A}^{\nu}L_{CA}\tilde{L}\tilde{L}_{CA}\boldsymbol{\sigma}_{\nu}%
^{C}\gamma_{0}^{C}L_{CA}L\tilde{L}_{CA}\gamma_{0}^{C}\nonumber\\
&  =r_{A}^{\nu}\tilde{L}^{\prime}\boldsymbol{\sigma}_{\nu}^{C}\tilde
{L}^{\prime\dag C}, \label{CarolP}%
\end{align}
where $L^{\prime}=L_{CA}L\tilde{L}_{CA}$ is the transformation from Alice to
Bob as seen by Carol \cite{Bay99}. In the special case that Carol shares
Alice's inertial frame, $L_{CA}=1,$ whereas if Carol and Bob share the same
frame, $L_{CA}=L.$

The key point is that the transformation (\ref{CarolP}) of measurables has
exactly the same form as the passive transformation in (\ref{transf2}). The
same result can be readily verified for the other transformations
(\ref{transf2}) of measurables. The transformations are the same no matter
which inertial observer is used for the mapping of APS to STA$^{+}$ and for
the definition of hermitian conjugation in STA$^{+}.$ Although the relation of
\emph{absolute APS} to STA$^{+}$ requires separate mappings and a separate
hermitian conjugation for each inertial observer, this may be considered an
artifact of the absolute approach in STA. Within \emph{relative APS }it makes
no difference which inertial observer is chosen; only one proper conjugation
and one proper (relative) basis $\left\{  \mathbf{e}_{\mu}\right\}  $ are
needed, and these can be associated with any inertial observer. The result can
also be described as choosing one absolute frame in STA as the \emph{observer
frame}. Since all inertial frames are equivalent, any inertial observer can be
assigned to this frame. Relative APS thus incorporates the basic principle of
relativity that all inertial observers are equivalent and that only the
relative motion and orientation of frames is physically significant.

When we organize the Lorentz transformations as in (\ref{transf2}) to compute
the measurables, such as the scalar components of a paravector $r_{A}$, we
find the transformations of a \emph{spacetime vector,} even in the special
case when $r_{A}=\mathbf{e}_{\mu}.$ We need to understand what it means for
$\mathbf{e}_{\mu}=\boldsymbol{\sigma}_{\mu}^{A}=\gamma_{\mu}^{A}\gamma_{0}%
^{A},$ a bivector in STA$^{+},$ to transform as a spacetime vector rather than
as a bivector as in (\ref{sigmatransf}). The paravector transformation of
$\mathbf{e}_{\mu}$ can be expressed in STA as%
\begin{equation}
\mathbf{e}_{\mu}\rightarrow L\mathbf{e}_{\mu}L^{\dag A}=L\gamma_{\mu}%
^{A}\gamma_{0}^{A}L^{\dag A}=\left(  L\gamma_{\mu}^{A}\tilde{L}\right)
\gamma_{0}^{A}~, \label{pvtransf}%
\end{equation}
where we have employed the definition (\ref{STAdag}) of Alice's proper
conjugation. The paravector basis elements $\mathbf{e}_{\mu}$ are defined to
be even elements of STA, and each involves the product of vectors from two
absolute STA frames. The paravector transformation (\ref{pvtransf}) transforms
only one of the frames. The result helps to clarify the roles of the two
factors: the first factor $\gamma_{\mu}^{A}$ gives the STA frame in which the
observed object property is expanded, and the second factor $\gamma_{0}^{A}$
gives the absolute proper velocity of the observer. The two factors generally
transform differently, as in the paravector transformation (\ref{pvtransf}).
The product of the two frame vectors gives the first relative to the second.
An observer will normally choose a proper basis in which the object and
observer frames are the same, but this is not necessary; she may also choose
an object frame in relative motion. In all cases, in the relative version of
APS the paravector basis elements are relative to the observer.

To obtain the transformation for spacetime bivectors from that for
paravectors, (\ref{pvtransf}), we form a product of two paravectors and
transform it:%
\[
\mathbf{e}_{\mu}=\mathbf{e}_{\mu}\mathbf{\bar{e}}_{0}\rightarrow
L\mathbf{e}_{\mu}L^{\dag}\overline{\left(  L\mathbf{e}_{0}L^{\dag}\right)
}=L\mathbf{e}_{\mu}\mathbf{\bar{e}}_{0}\bar{L}~.
\]
This is equivalent to the spacetime bivector transformation (\ref{sigmatransf}%
) of $\mathbf{\sigma}_{\mu}^{A}~.$ Of course $\mathbf{e}_{0}=1,$ but it is
often convenient to add factors of $\mathbf{e}_{0}$ or $\mathbf{\bar{e}}_{0}$
to display the correct covariant behavior.

To summarize, in relative APS, spacetime vectors are covariantly represented
by real paravectors. For example, the energy-momentum paravector of a particle
is $p=mu=E+\mathbf{p~.}$ Multiparavectors of higher grade can be formed to
represent other covariant geometrical objects, namely spacetime planes,
hypersurfaces, and volumes. Simple Lorentz rotations of paravectors have the
form (\ref{Lrotp}) where $L$ is the exponential of a biparavector representing
the plane of rotation. Since as discussed above, the spacetime vectors
represented by paravectors in APS are all \emph{relative} to the observer, a
single Lorentz rotation can equally well represent an active transformation of
the observed spacetime vector, the inverse passive transformation of the
observer, or some combination thereof. In the passive case, the spacetime
vector $p$ can be treated as invariant with respect to one observer, say
Alice, and the transformation derived by expressing Bob's frame relative to
Alice. Thus%
\[
p_{A}=p_{A}^{\mu}\mathbf{e}_{\mu}=p_{B}^{\nu}\mathbf{u}_{\nu}~,
\]
where $\mathbf{u}_{\mu}=L\mathbf{e}_{\mu}L^{\dag}$ is Bob's frame as seen by
Alice. This gives $p_{B}^{\nu}=p_{A}^{\mu}\left\langle \mathbf{e}_{\mu
}L\mathbf{\bar{e}}^{\nu}L^{\dag}\right\rangle _{S},$ which is exactly the same
relation as found from the transformation (\ref{Lrotp}).

Multiparavectors of grades 0 through 4 form linear subspaces of the algebra.
Grade-0 paravectors are the same real scalars as grade-0 vectors, but
covariantly they model spacetime scalars, which are invariant under Lorentz
transformations. Grade-1 paravectors model spacetime vectors and form the
four-dimensional paravector space that is also the direct sum of scalar and
vector spaces. The biparavectors model spacetime planes and form a
six-dimensional subspace from the direct sum of vector and bivector spaces.
The four-dimensional subspace of triparavectors, which model hypersurfaces in
spacetime, is the direct sum of bivector and trivector spaces. Finally, the
paravector volume element, which models the spacetime pseudoscalar, also
serves as the vector volume element:%
\[
\mathbf{e}_{0}\mathbf{\bar{e}}_{1}\mathbf{e}_{2}\mathbf{\bar{e}}%
_{3}\mathbf{=e}_{1}\mathbf{e}_{2}\mathbf{e}_{3}=i~.
\]
The split of any paravector, biparavector or triparavector into its
multivector parts is a space/time split in APS.

\section{Discussion}

The measurement of a physical vector or multivector depends both on the
vectors of the physical system to be measured and on the commoving frame of
the observer. STA represents both the object to be measured and the observer
in terms of abstract absolute frames, but the \emph{measurables }are
components of the object vectors on the basis vectors of the observer frame.
These measurables appear as even elements of STA, involving products of
vectors of the observed system with the frame vectors of the observer. If we
consider only a single observer, we can map the elements of STA$^{+}$ onto
those of APS in such a way that the basis elements $\mathbf{e}_{\mu}$ of APS
are just proper relative basis elements $\boldsymbol{\sigma}_{\mu}=\gamma
_{\mu}\gamma_{0}$ of the observer. Persistent spatial vectors of APS then
correspond to timelike bivectors of STA, and hermitian conjugation, which
gives reversion in APS, is defined using the proper velocity of the observer.
The result is that paravectors in APS are always given relative to the
observer. Only such relative paravectors and their products can be measured,
and this is precisely what is computed in APS. The additional flexibility
provided by STA, namely to compute absolute objects and observer frames
independently, although at times conceptually appealing, is never needed in
physical measurements. This is the reason that APS can represent any physical
process as well as STA.

When there are two or more observers, they generally determine distinct values
of the measurables, and an important goal of any relativistic formalism is to
relate such values. In the absolute version of APS, the proper paravector
basis $\left\{  \mathbf{e}_{\mu}\right\}  $, as well as the proper hermitian
conjugation, depends on the observer and is identified with an absolute frame
in STA$^{+}$. It is the observer's own proper conjugation operator that tells
her which parts of spacetime bivectors are her vectors and which are her
bivectors. Lorentz boosts between different observers mix spatial vectors and
bivectors, all of which lie in the six-dimensional space of spacetime
bivectors of STA$^{+}$. Such mixing explains why boosts of persistent real
vectors, such as the electric field, induce an observed magnetic field. A
similar transformation applies to each observer's proper basis vectors when
these are treated as persistent vectors. While the observer dependence of
hermitian conjugation means that each observer sees her own basis vectors as
real, she may find another observer's persistent basis vectors to be complex.

However, a different transformation applies to spacetime vectors such as
acceleration. There the separate conjugations applied by the two observers
ensure that the pure spatial vector seen by one observer remains real for the
other and that a scalar part is generated. Thus, two distinct boost
transformations for vectors ensue, one for persistent vectors such as the
electric field giving a complex vector, and the other for instantaneous
spacetime vectors such as the proper acceleration giving a real paravector.
The covariant nature of the object must be recognized in order to know which
transformation to apply.

To establish explicit transformations of the measurable coefficients, we can
write the transformations between observers in terms of a proper basis of a
single observer. We then find that the transformations do not depend on which
observer's proper basis and conjugation are used. A single proper basis and
conjugation (reversion) operation in APS can in fact be used to relate all
inertial observers, even though they have different absolute frames in STA.
This leads to the relative APS approach, in which all inertial observers share
a single conjugation and a single proper basis. It is consistent with the
principle of relativity that all inertial frames are equivalent and only
relative motion and orientation of frames matters. The different covariant
behavior of the proper acceleration and the electric field corresponds to the
difference between spacetime vectors and spacetime bivectors, and in relative
APS these are represented by real paravectors and biparavectors, respectively.
APS can be interpreted both in terms of spatial vectors and their products,
and in covariant terms, with Minkowski vectors of spacetime. The absolute
version of APS emphasizes the first and the relative version the second. In
both cases, one clearly identifies the elements of APS with measurables in
STA$^{+}$, and although in the presence of more than one observer the two
versions invoke different mappings to STA$^{+},$ we have shown them both to be
two different approaches within a single coherent geometric algebra, namely APS.

The ability of APS to model relativistic processes as faithfully as STA, even
though as a vector space it has only half the number of independent elements,
is due mainly to the additional structure required by the absolute-frame
representation of relativity in STA. APS, in contrast, does not model
non-observable absolute frames but concentrates instead on \emph{measurable}
properties relative\emph{ }to an observer. A characteristic of the APS
approach that also contributes to its efficiency is the double role of vector
grades. Thus, a scalar might be a Lorentz invariant or the time component of a
paravector, and a vector might be part of a paravector or part of a
biparavector. In this regard, APS models relativistic quantities as humans
commonly do. For example, the mass of a particle is both the Lorentz invariant
length of its momentum and the time component (in units with $c=1$) of the
momentum in its rest frame. This is naturally expressed in APS, where
$m=m\mathbf{e}_{0}$ is the same object in both roles, whereas in STA the roles
are played by distinct elements: $m\neq m\gamma_{0}~.$ This feature may be
considered an extension of a basic attraction of working with geometric
algebras instead of with separate vector spaces: in the algebra there is one
zero element, and one does not need to distinguish the zero scalar from the
zero vector or zero bivector.

\subsection*{Acknowledgment} One of us (WEB) thanks the Natural
Sciences and Engineering Research Council of Canada for support of
research.


\begin{thebibliography}{99}                                                                                               %


\bibitem {Hes66}D. Hestenes, \emph{Spacetime Algebra}, Gordon and Breach, New
York 1966.

\bibitem {Hes84}D. Hestenes and G. Sobczyk, \emph{Clifford Algebra to
Geometric Calculus: A Unified Language for Mathematics and Physics}, D.
Reidel, Dordrecht, 1984.

\bibitem {Hes87}D. Hestenes, \emph{New Foundations for Classical Mechanics,
}2nd edn., Kluwer Academic, Dordrecht, 1999.

\bibitem {Hes03}D. Hestenes, \emph{Am. J. Phys.} \textbf{71}:104--121, 2003;
\textbf{71}:691--714, 2003.

\bibitem {Bay96}W. E. Baylis, editor, \emph{Clifford (Geometric) Algebra with
Applications to Physics, Mathematics, and Engineering}, Birkh\"{a}user, Boston 1996.

\bibitem {Loun97}P. Lounesto, \emph{Clifford Algebras and Spinors}, second
edition, Cambridge University Press, Cambridge (UK) 2001.

\bibitem {Ablam03}R. Ab\l amowicz and G. Sobczyk, eds., \emph{Lectures on
Clifford Geometric Algebras,} Birkh\"{a}user, Boston, 2003.

\bibitem {Snygg97}J. Snygg, \emph{Clifford Algebra, a Computational Tool for
Physicists,} Oxford U. Press, Oxford, 1997.

\bibitem {GuSp97}K. G\"{u}rlebeck and W. Spr\"{o}ssig, \emph{Quaternions and
Clifford Calculus for Physicists and Engineers,} J. Wiley and Sons, New York, 1997.

\bibitem {Dor03}C. Doran and A. Lasenby, \emph{Geometric Algebra for
Physicists,} Cambridge University Press, Cambridge (UK), 2003.

\bibitem {Hes74}D. Hestenes, \emph{J. Math. Phys.}\textbf{ 15}, 1768--1777 (1974).

\bibitem {Bay99}W. E. Baylis, \emph{Electrodynamics: A Modern Geometric
Approach}, Birkh\"{a}user, Boston, 1999.

\bibitem {Bay89}W. E. Baylis and G. Jones, \emph{J. Phys. A (Math Gen)}
\textbf{22}, 1--16; 17--29 (1989).

\bibitem {Bay80}W. E. Baylis, \emph{Am. J. Phys.} \textbf{48}, 918--925 (1980).

\bibitem {Sob81}G. Sobczyk, \emph{Phys. Letters A} :45--48 Jul (1981).

\bibitem {Silberstein1912}L. Silberstein, Phil. Mag. \textbf{23}, 790--809 (1912).

\bibitem {Conway1912}A. W. Conway, Phil. Mag. \textbf{24}, 208 (1912).

\bibitem {Sob81a}G. Sobczyk, \emph{Acta Phys. Pol.}, Vol.\textbf{B12}, 509-521 (1981).
\end{thebibliography}
\end{document}